\documentclass[twoside,a4paper]{article}
\usepackage{graphicx}
\usepackage{amsmath}
\usepackage{amsthm}
\usepackage{multirow}

\theoremstyle{remark}

\theoremstyle{example}
\newtheorem{example}{Example}

\begin{document}
\title{Entropy Assessment of Windows OS Performance Counters}
\author{Richard Ostert\' ag\hskip8ex Martin Stanek\\[0.5em]
\normalsize\texttt{[ostertag, stanek]@dcs.fmph.uniba.sk}\\[0.5em]
\normalsize Department of Computer Science\\[-0.1em]
\normalsize Comenius University, Mlynsk\' a dolina\\[-0.1em]
\normalsize 842 48 Bratislava, Slovakia}
\date{}

\maketitle

\begin{abstract}
The security of many cryptographic constructions depends on random number generators for providing unpredictable keys, nonces, initialization vectors and other parameters. Modern operating systems implement cryptographic pseudo-random number generators (PRNGs) to fulfill this need. Performance counters and other system parameters are often used as a low-entropy source to initialize (seed) the generators. We perform an experiment to analyze all performance counters in standard installation of Microsoft Windows~7 operating system, and assess their suitability as entropy sources. Besides selecting top 19 counters, we analyze their mutual information (independence) as well as robustness in the virtual environment. Final selection contains 14 counters with sufficient overall entropy for practical applications.
\end{abstract}

\hspace{0.87cm}\textbf{Keywords:} Entropy, Windows performance counters, Randomness, PRNG.

\section{Introduction}

Random number generators are an important element in various cryptographic constructions. They are needed for generating keys, initialization vectors, and other parameters. The lack of randomness has devastating consequences for the security of cryptographic constructions and protocols. Well known examples are the broken seeding of openssl pseudo-random generator in Debian (resulting in limited set of potential keys generated by openssl), fixed parameter used in Sony PlayStation 3 implementation of ECDSA signing (resulting in compromised private key), and many other.
There are two types of random number generators \cite{NIST800-22}:
\begin{itemize}
\item (non-deterministic/true) random number generators use a non-deterministic entropy source together with some additional post-processing to produce randomness;
\item pseudo-random number generators --- typically deterministic algorithms that produce pseudo-random data from a seed (hence the seed must be random and unpredictable, produced by non-deterministic random number generator).
\end{itemize}

Various pseudo-random generators are standardized. For example, the list of approved random number generators for FIPS 140-2 can be found in \cite{FIPS140-2}. Modern operating systems implement random generators and allow application to use them. Linux uses CRC-like mixing of data from entropy sources to an entropy pool and extracts random data via SHA-1 from the entropy pool \cite{linux}, FreeBSD uses a variant of Yarrow algorithm \cite{freebsd}, Windows provides multiple PRNGs implementations via CNG (Cryptography, Next Generation) API \cite{win-secpol} with underlying generator compliant with NIST SP 800-90 \cite{NIST800-90A}. New generation of processors often contain hardware random numbers generators suitable for cryptographic purposes. 

The unpredictability of random number generators is essential for the security of cryptographic constructions. Good statistical properties, assessed by batteries of test such as NIST \cite{NIST800-22}, TestU01 \cite{TestU01} or DIEHARD \cite{diehard}, are necessary but not sufficient for unpredictability of the generator. An interesting assessment and details of Intel's Ivy Bridge hardware random number generator was published recently \cite{HKM2012}.

Since pseudo-random generators require seeding, usually multiple low-entropy sources are used to gather sufficient entropy into so-called entropy-pool for seeding or reseeding. Low-entropy sources are various system counters and variables, such as CPU performance and utilization characteristics, the usage of physical and virtual memory, file cache state, time, process and thread information, etc. An extensive list of operating system variables used to fill the Windows entropy pool can be found in \cite{win-cng}. An example of relatively simple design of PRNG based on timings from a hard disk drive is described in \cite{GKD04}.

\subsubsection*{Our contribution}

There can be various motivations for designing and implementing own pseudo-random number generator -- performance, a lack of trust to generators provided by operating system or other components (a well known story of Dual\_EC\_DRBG) or even those provided by CPU hardware \cite{BRPB13}, experimenting, additional features/properties not present in available generators, etc. All pseudo-random generators must be seeded. In case of Windows operating system we can easily use various counters as low-entropy sources.

We analyze the standard set of performance counters in Windows operating system. Our goal is to answer the following questions:

\begin{itemize}
\item What counters are best suited as entropy sources, and how much entropy they provide?
\item Are these sources independent or correlated?
\item Are these sources sufficiently independent on the operating system state (e.g. reboot or restoring from snapshot of a virtual system)?
\end{itemize}

Let us remind that counters are not used directly, they serve for seeding a pseudo-random generator. We do not expect the counters will satisfy some battery of statistical tests -- we measure primarily the entropy of individual counters and the mutual information of promising counters.

Section \ref{prelim} defines some notions and it also justifies a method of preprocessing the counters. Results of our experiments are presented in Section \ref{result}. We summarize the implications of the results and outline the possible further research in Section \ref{concl}.

\section{Preliminaries} 
\label{prelim}

\subsection{Entropy measures}

Let $X$ be a discrete random variable with values $\{x_1,\dots, x_n\}$. Let $p_i$ denote the probability that $X$ attains the value $x_i$, i.e. $\Pr[X=x_i]=p_i$. A standard measure of uncertainty is the Shannon entropy of a random variable. We denote it $H_1(X)$:
\[
H_1(X) = -\sum_{i=1}^n p_i \lg p_i,
\]
where $\lg$ denotes a logarithm with base 2, and we use a convention $0\cdot \lg 0 = 0$ when necessary. The smallest entropy in the class of all R\'enyi entropies \cite{R61} is a min-entropy $H_\infty(X)$:
\[
H_\infty(X) = -\lg \max_{1\leq i\leq n} p_i.
\]
Particularly, $H_\infty(X)\leq H_1(X)$ for any arbitrary discrete random variable $X$. The min-entropy measures the uncertainty of guessing the value of $X$ in a single attempt.

It is impossible to know exact probability distributions for most of the operating system counters. We sample the values of the counters in our experiments. After ``preprocessing'' (see Section \ref{ppcnt}) we use a maximum likelihood estimate, where each probability is estimated as ratio of observed values to the number of samples. There are better estimators of entropy with reduced bias, such as Miller-Madow, NSB and others \cite{NSB2002,Pa2003}. However, we think that simple estimate is sufficient for our classification of the performance counters.

Let $Z=z_1,\dots,z_N$ be a sequence of values from a finite set $S$. According to the previous paragraph, we use notations $H_1(Z)$ and $H_\infty(Z)$ for the entropy and the min-entropy, respectively, where the probabilities are estimated as follows: $p_s = |\{z_i ;\; z_i = s,\; 1\leq i\leq N\}|/N$ for all $s\in S$.

\subsection{Preprocessing of counters}
\label{ppcnt}

There are two inherent problems when measuring the randomness of operating system counters. The first problem is that some counters, in fact, ``count''. Therefore, we get a sequence of increasing values depending on chosen sampling interval. Calculating entropy based on unique values (regardless of their non/uniformity) yields always the same value -- for $N$ samples we get $H_1(X)=H_\infty(X)=\lg N$. In order to have meaningful entropy values, we have to limit the possible values/outcomes of the random variable (for example by splitting counter into smaller chunks). The second problem is that counters values are presented as 64-bit integers and therefore the majority of bits are constant for any reasonable amount of time. This can distort the measured entropies and lead to unrealistic low estimates, when we use all bits of a counter. 

\begin{example}
Let us assume a fictitious 64-bit counter $C_1$, where the last bit of $C_1$ is random and all other bits of $C_1$ are zero, i.e. counter can have only two different 64-bit values $(0\dots00)_2$ and $(0\dots01)_2$. It is easy to see that $H_1(C_1)=1$.
However, splitting $C_1$ into bits and calculating the entropy from each bit of the counter yields: $H_1(1/128; 127/128)\approx 0.07$ (if we divide the above mentioned two 64-bit values to bits we get 128 bits in total, where 127 are equal to zero and only one is non-zero).
Splitting $C_1$ into bytes yields: $H_1(1/16; 15/16)\approx 0.34$ (we get 16 bytes, where one is equal to $(00000001)_2$ and remaining 15 bytes are equal to zero).
\end{example}

We deal with these problems by preprocessing the values by a transformation $T_\alpha$. Let $V=(v_{t-1}\ldots v_1v_0)$ be a $t$-bit vector. Let $\alpha$ be a positive integer such that $\alpha$ divides $t$. The transformation $T_\alpha$ produces $\alpha$-bit output from $t$-bit input:

\[
T_\alpha(V) = (v_{t-1}\ldots v_{t-\alpha}) \oplus (v_{t-\alpha-1}\ldots v_{t-2\alpha}) 
\oplus \ldots \oplus (v_{\alpha-1}\ldots v_{0})
\]

The transformation is chosen with the aim of simplicity. Notice that a complex transformation can distort the estimates in the opposite direction -- even simple incremental counter transformed by a cryptographically strong hash function will look like ideal random source.

The parameter $\alpha$ can be interpreted as an appetite for getting as much randomness as possible from a single value. Value of $\alpha$ affects the estimated entropy. As noted earlier in the discussion on random counters vs. simple incremental counters, too high $\alpha$ results in overestimating the entropy. On the other hand, too small $\alpha$ can be very pessimistic -- for example, we lose $t-\alpha$ bits of random data for each value of a truly random counter. 

In order to deal with counters that increment regularly we apply difference operator to the sampled counter values $(c_0,c_1,\dots,c_N)$:
\[
\triangle (c_0,c_1,c_2,\dots,c_{N-1}c_N) = (c_1-c_0,c_2-c_1,\dots, c_N-c_{N-1}).
\]
Trivially, the difference of random, independent values yields again random and independent values. Applying the difference operator allows measuring the entropy of change between successive samples of the counter. Since the difference operator does not do any harm to ``good'' counters we apply it to all counters. Moreover, after applying $T_\alpha \circ \triangle$ (i.e. $\triangle$ is applied first and $T_\alpha$ second) we group the obtained values into bytes. We take bytes as values of the random variable we use to measure the entropy. 

To represent negative numbers, introduced by the difference operator, we use their absolute value with a sign at the most significant bit position.

\section{Experiments}
\label{result}

A testing environment consisted of virtual PC emulated in VMware Workstation. The virtual PC was created with 2 vCPU, 2\,GB RAM and default settings of virtualization platform. We used clean installation of Windows 7, the desktop operating system with the largest share world-wide \cite{OSshare13}. We implemented our experiments using .NET platform. We also applied SP1 and all available updates (hot-fixes) provided by Microsoft. During experiments the virtual PC was idle, with no user interaction, i.e. only the sampling application and default services were running.

\subsection{Eliminating counters}

Our first step was to eliminate weak counters and select promising ones. We proceed as follows:
\begin{enumerate}
\item Enumerate all operating system's performance counters. We identified 1367 counters.
\item Let us define a sampling round as follows: sample all counters and then wait for 20\,ms. 
\item Repeat the sampling round to collect 10\,000 values for each counter. Eliminate counters that are constant, i.e. the entropy is zero. This resulted in 273 counters.
\item Re-sample remaining 273 counters in 100\,001 sampling rounds. Eliminate counters that are constant -- we got 266 counters after this elimination.

It is interesting that we have got counters non-constant in 10\,000 samples, but constant in 100\,001 samples. This anomaly was caused by some unknown rare event which happened while sampling the first 10\,000 samples and then never repeated during the first sampling and even the second, ten times longer, sampling period. This event influenced the following six counters (they were constant before and after this event):
\begin{itemize}
\item ``Cache, Read Aheads/sec" --- increased
\item ``Event Tracing for Windows, Total Number of Distinct Enabled Providers" --- decreased
\item ``Event Tracing for Windows, Total Number of Distinct Pre-Enabled Providers" --- increased
\item ``LogicalDisk, Split IO/Sec, \_Total" --- increased
\item ``PhysicalDisk, Split IO/Sec, \_Total" --- increased
\item ``Synchronization, Exec. Resource no-Waits AcqExclLite/sec, \_Total" --- increased
\end{itemize}
The last ``strange'' counter was ``Memory, System Code Resident Bytes" (see Figure \ref{figC}). This counter was oscillating in range from 2\,527\,232 to 2\,576\,384 for first 1\,643 samples and then settled down to the maximal value of the range (2\,576\,384 bytes $\equiv$ 629 pages of 4\,kB) and never changed again.

\begin{figure}[h]
\hfil
\includegraphics[scale=0.75]{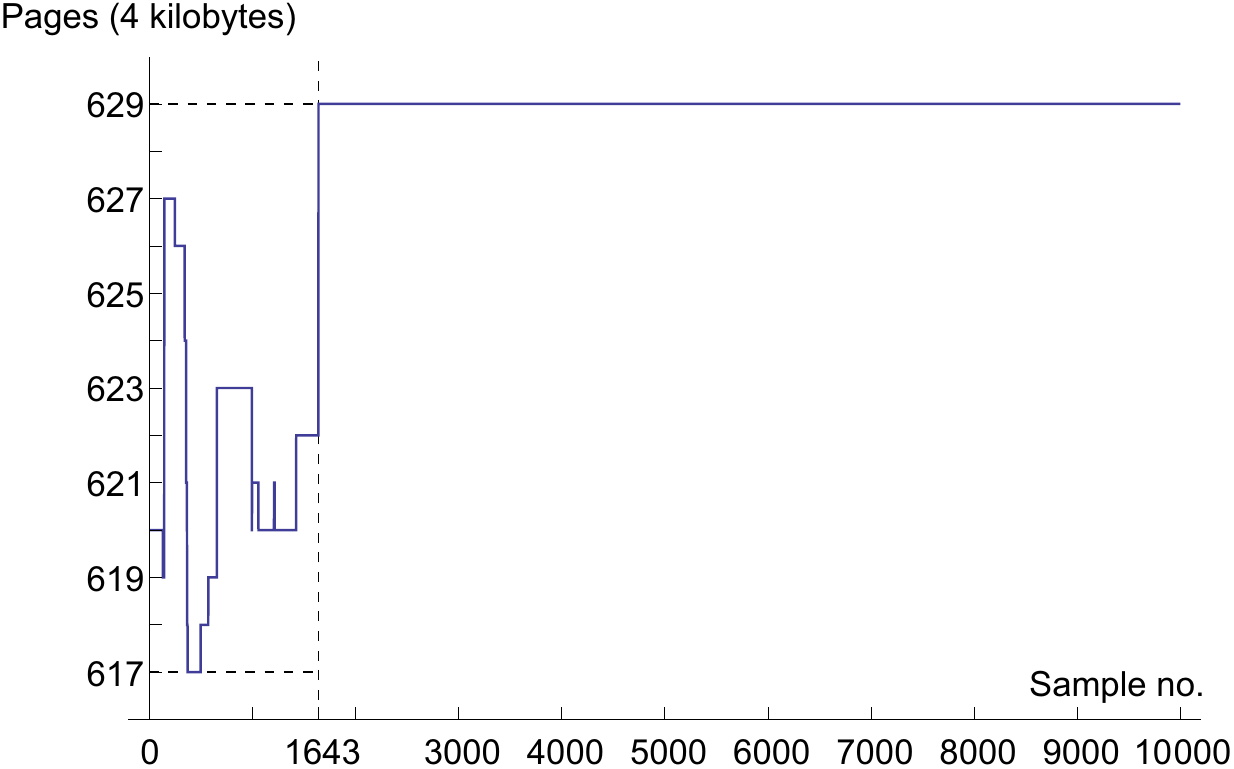}
\caption{``Memory, System Code Resident Bytes" oscillation for first 1\,643 samples.}\label{figC}
\end{figure}

\item Apply the difference operator to obtained values (now we have 100\,000 delta-values for each counter). After eliminating counters with constant delta-values, we obtained 263 counters.
\item Apply the transformation $T_\alpha$ for $\alpha=1,2,4,8$, and group individual results into bytes. Thus we get for each counter four separate sequences of bytes (12\,500, 25\,000, 50\,000 and 100\,000 bytes for $\alpha= 1,2,4,8$ respectively). We denote these sequences for counter $C$ as $C^{(1)}$, $C^{(2)}$, $C^{(4)}$ and $C^{(8)}$.

Setting $\alpha$ too high yields low entropy per bit values. For example when $\alpha=16$ no counter have value of $H_1+H_\infty>1.81$ per bit. For $\alpha=32$ no counter have value of $H_1+H_\infty>1$ per bit, i.e. no counters fill be in upper triangle  on Figure \ref{fig1}. Experiments imply that decreasing $\alpha$ generally increases scaled entropy value per bit.

\item Eliminate counters that are constant for some $\alpha$. None of the counters were eliminated in this step. We call all remaining 263 counters ``green''.
\end{enumerate}

The result of this process is summarized in Table \ref{tab1}.

\begin{table}[h]
\begin{center}
\begin{tabular}{lr}
Number of counters & 1367 \\
Non-constant counters after re-sampling & 266 \\
Non-constant counters after $\triangle$ & 263 \\
Non-constant counters after $T_\alpha$ & 263
\end{tabular}
\end{center}
\vspace{-1em}
\caption{Overview of counters elimination.}\label{tab1}
\end{table}

We computed the entropy and the min-entropy for all 263 remaining (green) counters. Since there are four sequences for each counter (for different $\alpha$), we took the minimum of their entropies (we believe that the counter's entropy should be sufficiently ``robust'' to simple transformations such as $T_\alpha$):
\begin{align*}
H_1(C) &= \min\{ H_1(C^{(1)}), H_1(C^{(2)}), H_1(C^{(4)}), H_1(C^{(8)})\} \\
H_\infty(C) &= \min\{ H_\infty(C^{(1)}), H_\infty(C^{(2)}), H_\infty(C^{(4)}), H_\infty(C^{(8)})\} \\
\end{align*}\vspace{-1cm}

Figure \ref{fig1} shows the distribution of $H_1$ and $H_\infty$ values for the green counters. The values of entropies are scaled per bit, i.e. divided by $8$.

\bigskip
\begin{figure}[h]
\hfil
\includegraphics[scale=0.75]{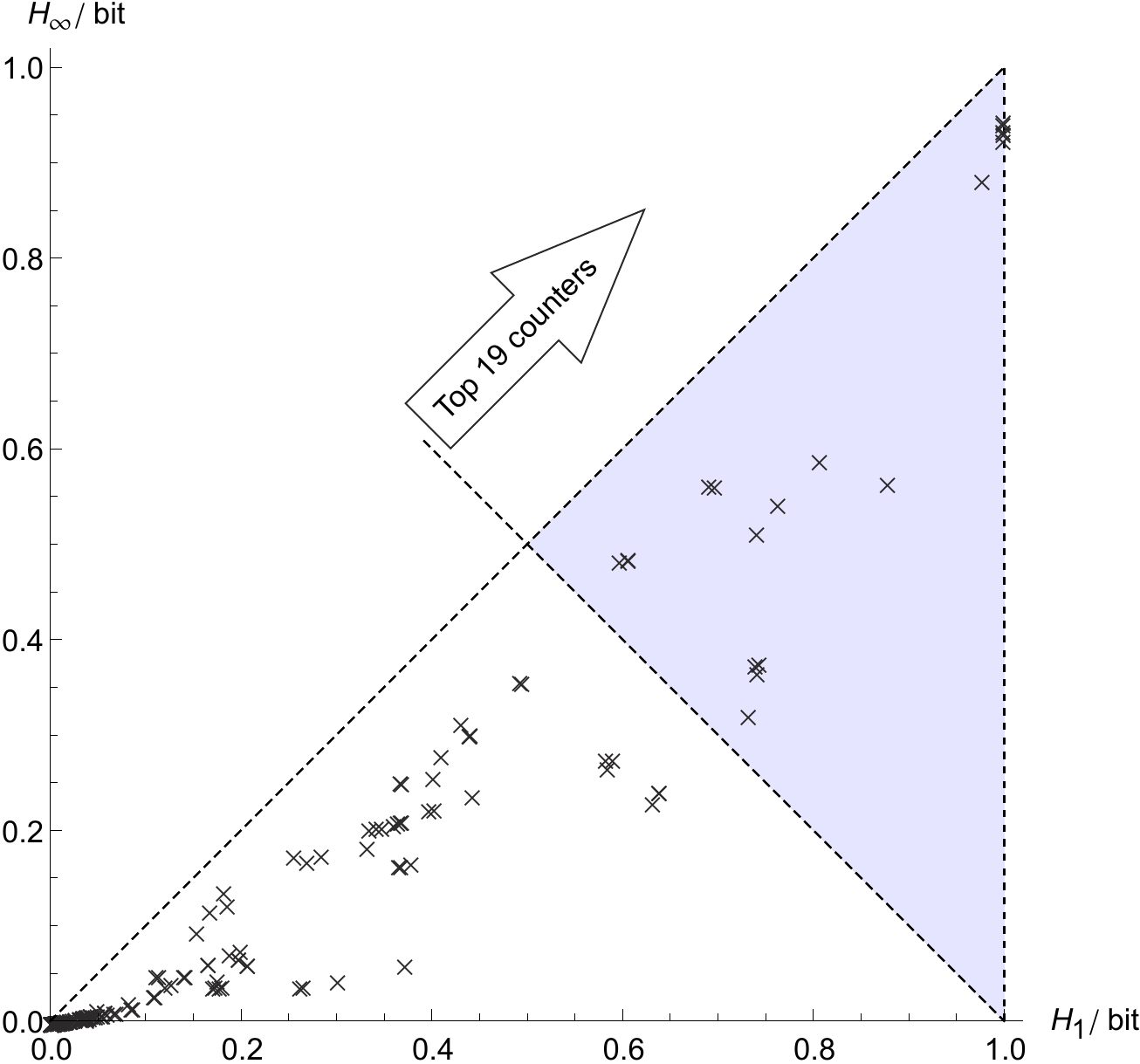}
\caption{Distribution of $H_1$ and $H_\infty$ for 263 green counters.}\label{fig1}
\end{figure}
\bigskip
\bigskip
\begin{figure}[h]
\hfil
\includegraphics[scale=0.75]{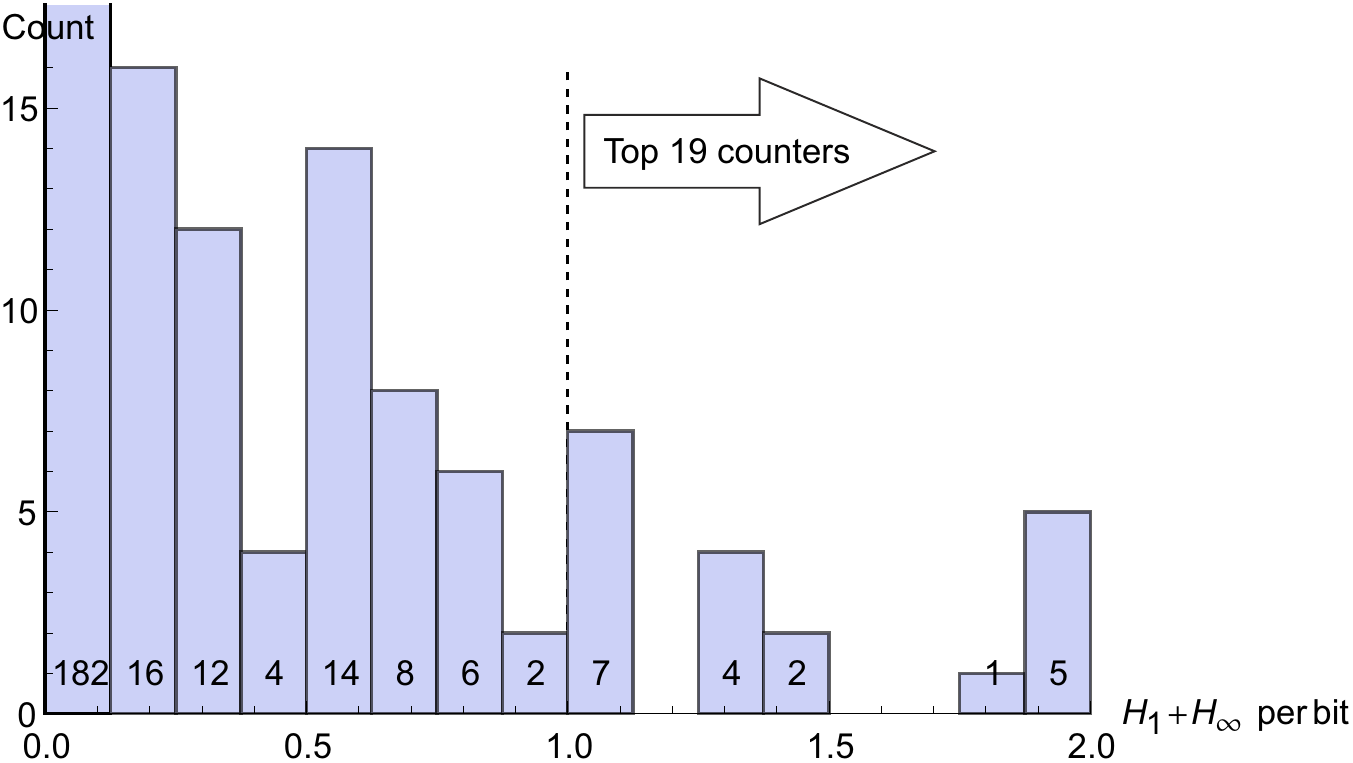}
\caption{Distribution of $H_1(C)+H_\infty(C)$ for 263 green counters.}\label{fig2}
\end{figure}

\subsubsection*{Top 19 counters}

A very defensive approach in using counters to form an entropy pool is to take all of them. With a suitable function to distill entropy from the pool we do not lose a single bit of entropy. However, periodically querying large number of counters costs additional CPU cycles and it might be unnecessary. We selected the most promising counters (according the highest value of $H_1(C)+H_\infty(C)$) for further analysis. Top 19 counters are summarized in Table \ref{tab2} (the entropies are scaled per bit, i.e. divided by 8).

\begin{table}[h]
\begin{center}
\begin{tabular}{cclccc}
\hline\noalign{\smallskip}
$C^{19}_i$ & $C^{14}_i$ & Counter description \hskip3.3cm per bit $\rightarrow$ & $H_1$ & $H_\infty$ & $H_1+H_\infty$ \\
\noalign{\smallskip}
\hline
\noalign{\smallskip}
1 & 1 & .NET CLR Memory, \% Time in GC & 0.9984 & 0.9457 & 1.9441 \\
2 & 2 & .NET CLR Security, \% Time in RT checks & 0.9981 & 0.9429 & 1.9410 \\
3 & 3 & Processor Information, \% C1 Time & 0.9982 & 0.9350 & 1.9332 \\
4 & 4 & .NET CLR Memory, Allocated Bytes/sec & 0.9984 & 0.9325 & 1.9309 \\
5 & 5 & Processor, \% C1 Time & 0.9981 & 0.9250 & 1.9231 \\
6 & 6 & Synchronization, Spinlock Acquires/sec & 0.9762 & 0.8828 & 1.8590 \\
7 & 7 & System, System Calls/sec & 0.8772 & 0.5654 & 1.4426 \\
8 & 8 & PhysicalDisk, \% Idle Time & 0.8058 & 0.5892 & 1.3950 \\
9 & 9 & Memory, Pool Nonpaged Allocs & 0.7620 & 0.5437 & 1.3057 \\
10 & 10 & Thread, Context Switches/sec & 0.6958 & 0.5626 & 1.2584 \\
11 & -- & System, Context Switches/sec & 0.6898 & 0.5638 & 1.2536 \\
12 & 11 & LogicalDisk, \% Idle Time & 0.7398 & 0.5135 & 1.2533 \\
13 & 12 & Memory, Available Bytes & 0.7424 & 0.3773 & 1.1197 \\
14 & -- & Memory, Free \& Zero Page List Bytes & 0.7382 & 0.3756 & 1.1138 \\
15 & -- & Memory, Available KBytes & 0.7403 & 0.3675 & 1.1078 \\
16 & 13 & Processor Information, C1 Transitions/sec & 0.6050 & 0.4869 & 1.0919 \\
17 & -- & Processor, C1 Transitions/sec & 0.6051 & 0.4858 & 1.0909 \\
18 & -- & Synchronization, IPI Send Software Interrupts/sec & 0.5959 & 0.4846 & 1.0805 \\
19 & 14 & Synchronization, Exec. Resource Total Acquires/sec & 0.7313 & 0.3222 & 1.0535 \\
\hline
\end{tabular}
\end{center}
\caption{Top 19 counters and their entropy and min-entropy per bit (sorted by $H_1+H_\infty$ per bit).}\label{tab2}
\end{table}

The counters are sorted according to the combined metric $H_1(C)+ H_\infty(C)$ in descending order. Interestingly but expectedly, there are three .NET counters among the best counters -- it is caused by the fact that .NET was the platform used for implementation of our experiment. 

\subsection{Mutual information of promising counters}
\label{mutual}

In an ideal world the counters would be mutually independent. However, some counters measure similar or related events, e.g. LogicalDisk and PhysicalDisk counters. In order to measure the mutual dependence of each pair of counters, we estimate the mutual information between all pairs of counters from our top 19 selection. 

The mutual information of two random variables $X$ and $Y$ is defined as follows: 
\[
I(X;Y) = H_1(X) + H_1(Y) - H_1(X,Y).
\]
There are also other equivalent means to express the mutual information, however we use this formula for our computations. The mutual information is nonnegative and it is zero if and only if the random variables are independent. Stated informally, the closer $I(X;Y)$ is to zero, the more independent are variables $X$ and $Y$.

Since the mutual information uses entropy $H_1(X,Y)$, we divided the transformed counter's values to half-bytes (from original bytes). This guarantees sufficient number of samples for estimating $H_1(X,Y)$. Therefore we have $0\leq I(C_i;C_j)\leq 4$ for any pair of counters -- with $0$ for independent and $4$ for functionally dependent pair.

\begin{figure}[h]
\hfil
\includegraphics[scale=0.85]{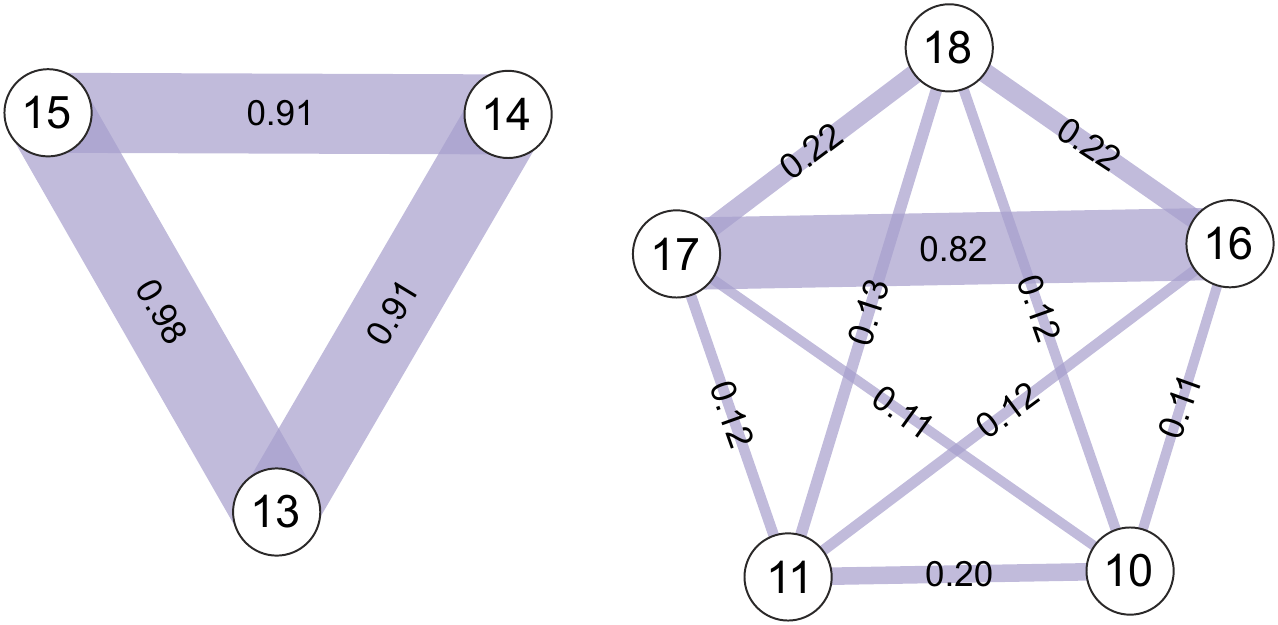}
\caption{Mutual information of dependent counters (values are divided by $4$). Independent counters (isolated vertices) are not shown.}\label{fig4}
\end{figure}

Most of the counters are mutually (almost) independent. Exceptions are groups of counters shown in Figure \ref{fig4} and Table \ref{tab3}. The first group of three dependent counters are memory counters for available memory and free \& zero page list with mutual information $I(C_i;C_j)/4 > 0.90$. The second group contains system and thread context switches counters, C1 transitions counters and one synchronization counter. This group can be further subdivided into two subsets, see Table \ref{tab3}. The experiment yielded $I(C_i;C_j)/4 < 0.10$ for all other pairs of counters, and we can consider them independent.

\begin{table}[h]
\begin{center}
\begin{tabular}{cclccc}
\hline\noalign{\smallskip\smallskip}
\multirow{1}{*}{$C^{19}_i$} & \multirow{1}{*}{$C^{14}_i$} & \multirow{1}{*}{Counter description} & $\displaystyle\frac{H_1+H_\infty}{8}$ & $\displaystyle\max_{i,j\text{ in group}} \frac{I(C^{19}_i;C^{19}_j)}{4}$ \\
\noalign{\smallskip}
\hline
\noalign{\smallskip}
\textbf{13} & \textbf{12} &\textbf{Memory, Available Bytes} & \textbf{1.1197} & \multirow{3}{*}{0.98} \\
14 & -- & Memory, Free \& Zero Page List Bytes & 1.1138 & \\
15 & -- & Memory, Available KBytes & 1.1078 & \\
\hline
\noalign{\smallskip}
\textbf{16} & \textbf{13} & \textbf{Processor Information, C1 Transitions/sec} & \textbf{1.0919} & \multirow{3}{*}{0.82} \\
17 & -- & Processor, C1 Transitions/sec & 1.0909 & \\
18 & -- & Synchronization, IPI Send Software Interrupts/sec & 1.0805 & \\
\hline
\noalign{\smallskip}
\textbf{10} & \textbf{10} & \textbf{Thread, Context Switches/sec} & \textbf{1.2584} & \multirow{2}{*}{0.20} \\
11 & -- & System, Context Switches/sec & 1.2536 & \\
\hline
\end{tabular}
\end{center}
\caption{Dependent groups with maximal mutual information in each group. Highlighted counters are selected as group representants for further analysis.}\label{tab3}
\end{table}

For further analysis we select all independent counters and one counter for each dependent group (the counter with highest value of our combined entropy metric). This reduces the number of counters to $14$, with total $H_1$ entropy $111.55$ bits per each 264\,ms -- this is a conservative estimate assuming $\alpha=1$ (time includes 20\,ms of sleeping and 13\,ms of collecting counters for each one of eight rounds).

\subsection{Robustness in virtual environment}

Virtual servers and PCs (VDI) are increasingly common in everyday reality of corporate IT. We try to answer a natural question about independence of selected counters in this environment. We created a snapshot of our (virtual) experimental Windows 7 system. After resuming the system, sampling starts immediately. We collect data for three resumptions from the same snapshot. Similarly to Section \ref{mutual} we compute mutual information, this time it is the mutual information between samples of the same counter. Moreover, in order to focus on potential correlations in short time span, the mutual information is calculated using floating window of length 1400 half-byte samples. This window corresponds to roughly 3 minutes, for sampling interval 20\,ms and $\alpha=1$.

Before diving into more detailed analysis, let us emphasize that the experiment showed mostly negligible mutual information of different runs of each counter. With exception of short spike of counter $C^{14}_{9}$ (Memory, Pool Nonpaged Allocs), all other mutual informations are way below 0.10, the value we used to declare independence of two counters in Section \ref{mutual}. Therefore, even starting sampling from the same snapshot will produce a sufficiently different content of entropy pool.

As we have three runs for every counter, we compute the mutual information between all three pairs of these runs. Then the minimum, the average (arithmetic mean) and the maximum of all obtained values are calculated. For comparison, we created three runs of 14 random, uniformly distributed and independent counters, and we computed their mutual information minimum, average and maximum.\footnote{Even if mutual information of independent random variables is zero, in finite number of samples we can expect some deviation from this theoretical value.} The comparison of real and random counters revealed that there are two types of counters. The first group contains counters with mutual information statistics similar to random counters -- these are counters $C^{14}_i$, for $i \in \{2, 3, 5, 6, 7, 8, 10, 11, 13, 14\}$ and we call them the upper group/counters. The second group of counters $C^{14}_i$, for $i \in \{1, 4, 9, 12\}$ exhibits statistics even better than the random counters, therefore we call them lower group/counters. Numerically, groups are divided by threshold value $0.021$. If mutual information between all pairs of runs of the counter lies above the threshold, then the counter belongs in the upper group, otherwise it is in the lower group.

\begin{figure}[h]
\hfil
\includegraphics[scale=0.79]{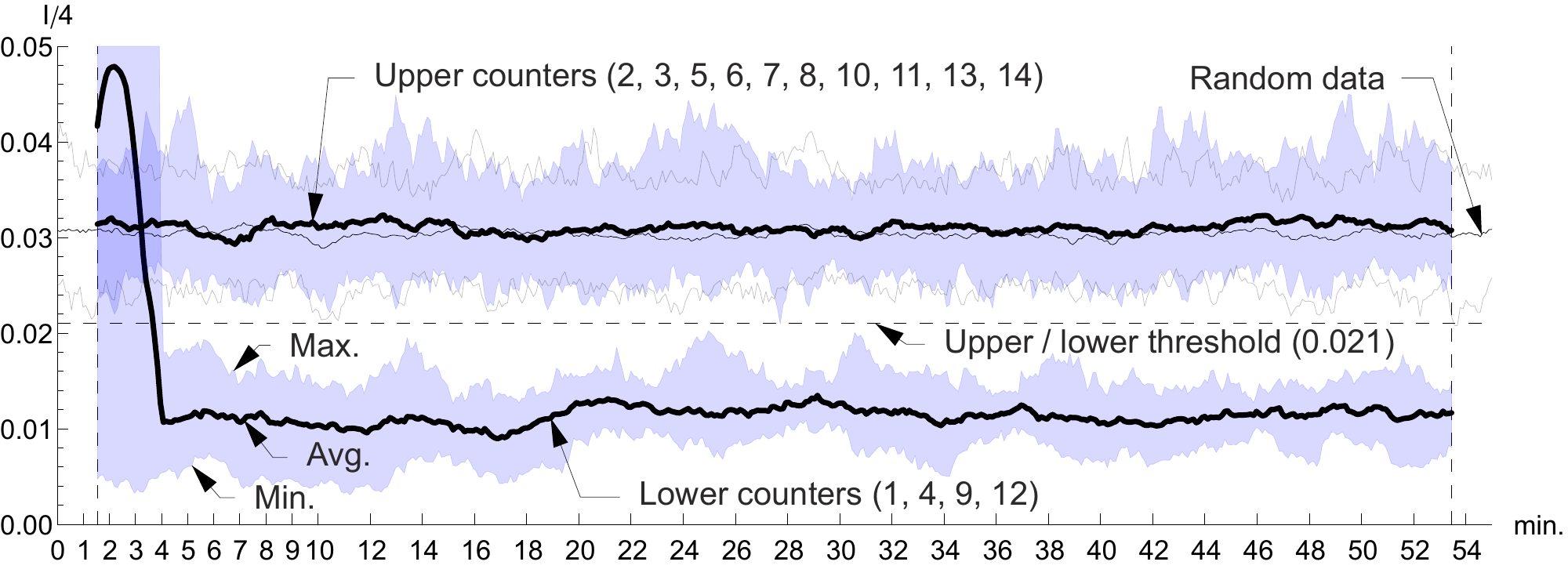}
\caption{
Mutual information between different runs, i.e. resumptions from the same snapshot. Values are divided by $4$ as marked on Y axis. Upper and lower group counters are presented separately.
}\label{fig5}
\end{figure}

The results for the selected 14 counters, divided into upper and lower groups, are presented in Figure \ref{fig5}.

\begin{figure}[h]
\hfil
\includegraphics[scale=0.79]{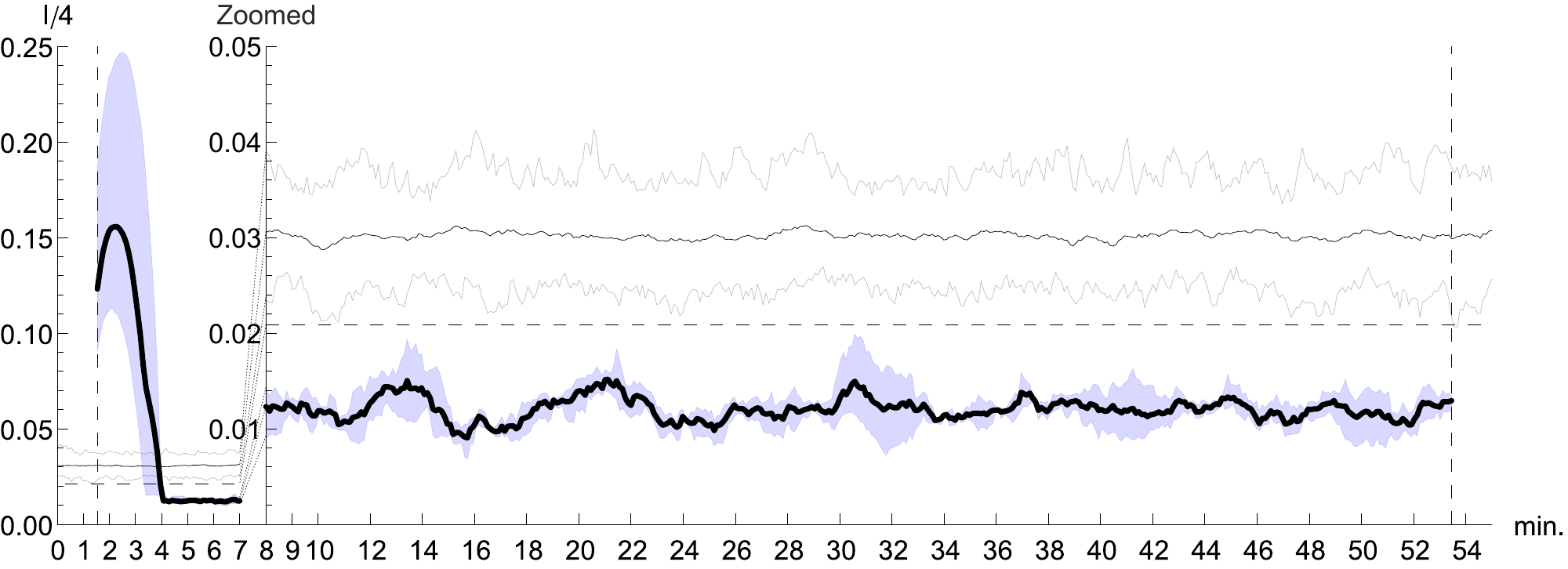}
\caption{Mutual information between different runs for counter $C^{14}_{9}$ (Memory, Pool Nonpaged Allocs).}\label{fig6}
\end{figure}

Unusual behavior of the lower counters at the beginning is caused by counter $C^{14}_{9}$ (Memory, Pool Nonpaged Allocs) and, to a lesser extent, counter $C^{14}_{12}$ (Memory, Available Bytes). Other two counter from the lower group show stable behavior on entire interval. Figure \ref{fig6} shows counter $C^{14}_{9}$ statistics -- we can observe a sudden increase of mutual information roughly after 2.5 minutes after resumption.

Another ``strange'' counter is counter $C^{14}_{12}$. The mutual information is higher than threshold in the first 3 minutes (but still less than average of random counters). After that, the values drop below threshold and stay there. 

\begin{figure}[h]
\hfil
\includegraphics[scale=0.79]{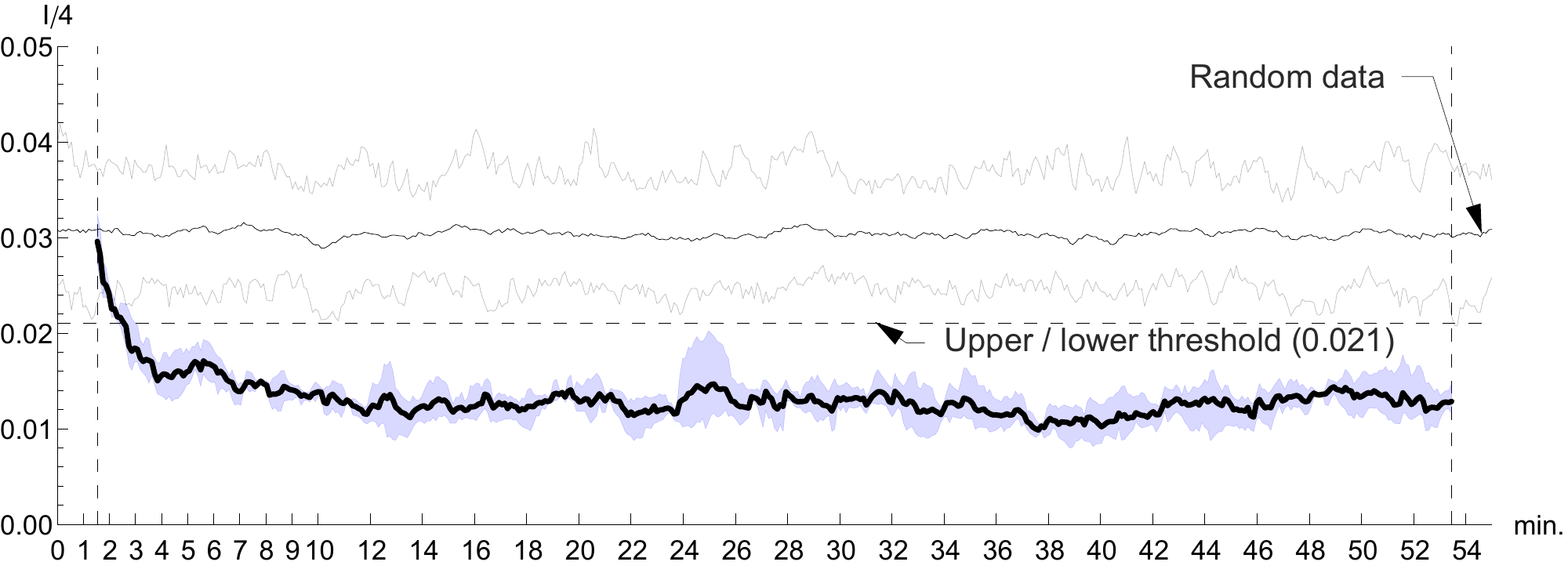}
\caption{Mutual information between different runs for counter $C^{14}_{12}$ (Memory, Available Bytes).}\label{fig7}
\end{figure}

The lower group has distinctively lower mutual information statistics than completely random counters. This makes 
these counters slightly ``artificial'' and probably suspicious.

\section{Conclusion}
\label{concl}

We analyzed Windows 7 performance counters as potential sources of randomness. Generally, more counters and other entropy sources you include into your entropy pool the better chance of having sufficient entropy when you need it. Our experiments yielded 19 promising counters. The final selection consists of 14 counters with enough entropy for practical purposes. Their analysis allows us to draw the following conclusions:
\begin{itemize}
\item If your applications uses .NET platform, some of the platform's performance counters are good entropy sources.
\item Using 11 counters (without .NET counter) yields $87.59$ bits per 240\,ms -- a conservative estimate assuming $\alpha=1$ (time includes 20\,ms of sleeping and 10\,ms of collecting counters for each one of eight rounds). Adding three .NET counters increases the entropy to $111.55$ bits per 264\,ms ($\alpha=1$, time includes 20\,ms of sleeping and 13\,ms of collecting counters for each one of eight rounds). Even without detailed analysis of unpredictability of these sources we can conclude that Windows performance counters are viable option to feed randomness pool for PRNGs.
\item Interestingly, most of top counters are mutually independent. A strong mutual dependence between counters is usually observed with obvious pairs (e.g. available memory in bytes and kilobytes).
\item Selected counters are robust entropy sources in virtual environment. Independent runs of the virtual PC from the same snapshot showed mutual independence of each counter's samples (with exception of short spike of counter $C^{14}_{9}$).\footnote{Although one might expect that independent runs of the virtual PC from the same snapshot will result in the same or in highly correlated performance counters evolution, the experiments shows the opposite. Apparently, there are so many events in a running operating system with complex relationships that the events appear as nondeterministic.}
\end{itemize}

We did our analysis only in a virtual environment. We expect that the experiment in a host environment (physical hardware) would show comparable or even better results. Certainly, the analysis can be extended further by more thorough experiments (e.g. considering various time interval for sampling, exploring mutual dependence of higher orders, changing preprocessing of sampled counters), using better entropy estimators (e.g. NSB estimator), studying the possibilities of influencing the predictability of performance counters etc.

\paragraph{Acknowledgment.} The authors acknowledge support by VEGA 1/0259/13.

\end{document}